\documentclass[10pt,showpacs,twocolumn,amsmath,amssymb,aps,pre,preprintnumbers]{revtex4-1}
\usepackage{graphicx,latexsym,amsfonts,dcolumn,bm,amsthm}
\newcommand{\kc}{k_{\textrm{\tiny C}}} \newcommand{\Nc}{N_{\textrm{\tiny C}}}
\newcommand{\kin}{k_{\textrm{\scriptsize in}}}
\newcommand{\pin}{p_{\textrm{\scriptsize in}}}
\newcommand{\pout}{p_{\textrm{\scriptsize out}}}
\newcommand{\kout}{k_{\textrm{\scriptsize out}}}
\newcommand{\nD}{n_\text{\scriptsize D}}
\newcommand{\ND}{N_\text{\scriptsize D}}
\begin{document}
\preprint{Published in PLoS One {\bf(8}):e82578, (2013). DOI: 10.1371/journal.pone.0082578}
\bibliographystyle{nature}
%
\title{Resilience and controllability of dynamic collective behaviors}
\author{Mohammad Komareji$^1$}
\author{Roland Bouffanais$^1$}%
\email{bouffanais@sutd.edu.sg}
\affiliation{$^1$Singapore University of Technology and Design, 20 Dover
  Drive, Singapore 138682}
\date{October 21, 2013}%
\begin{abstract}
The network paradigm is used to gain insight into the structural root causes
of the resilience of consensus in dynamic collective behaviors, and to analyze
the controllability of the swarm dynamics. Here we devise the dynamic
signaling network which is the information transfer channel underpinning the
swarm dynamics of the directed interagent connectivity based on a topological
neighborhood of interactions. The study of the connectedness of the swarm
signaling network reveals the profound relationship between group size and
number of interacting neighbors, which is found to be in good agreement with
field observations on flock of starlings [Ballerini \textit{et al.}  (2008)
Proc. Natl. Acad. Sci. USA, 105: 1232]. Using a dynamical model, we generate
dynamic collective behaviors enabling us to uncover that the swarm signaling
network is a homogeneous clustered small-world network, thus facilitating
emergent outcomes if connectedness is maintained. Resilience of the emergent
consensus is tested by introducing exogenous environmental noise, which
ultimately stresses how deeply intertwined are the swarm dynamics in the
physical and network spaces. The availability of the signaling network allows
us to analytically establish for the first time the number of driver agents
necessary to fully control the swarm dynamics.
\end{abstract}
\maketitle
\section*{Introduction}
%
In an animal group, if each individual contributes independently to a given
collective goal or objective, the resulting group behavior follows some sort
of normal distribution pattern. On the contrary, if animals work collectively
with a certain level of local interaction or communication, the output of
their acts is more than the sum of each individual's
act~\cite{viscek2012}. The emergent behavior is thus characterized by some
signatures in the structural properties of the network underpinning their
cooperative
behavior~\cite{viscek2012,barabasi,miller,bak,mitchell,cavagna:scale}. Moreover,
the global outcome of their local interactions heavily depends on each
individual's initial conditions~\cite{kattas,slotine}. For example the velocity
of a flock of birds was found to be a function of each bird's initial
velocity\cite{rezaflock}. The emergence of spatiotemporal order at the group
level has been observed in many biological systems~\cite{camazine}---insect
colonies, fish schooling, bird flocking, amoebae aggregating, bacteria
swarming, in many human
activities~\cite{helbing97:_model,nagel96:_partic}---pedestrian and automobile
traffic, and in the artificial world with robotic swarm
systems~\cite{hsieh08:_decen}.

Sumpter~\cite{sumpter} argues that the key to understanding collective
behaviors---and more broadly the concept of self-organization---lies in
identifying the principles of the behavioral algorithms followed by individual
animals and how information flows between the animals. That is what
physicists, biologists and engineers have been trying to achieve through
Lagrangian modeling of animals' collective behaviors as attested by the
significant body of literature dealing with this specific
issue~\cite{viscek2012,reynolds,vicsek,jad,couzin,rezaflock,naomi,cucker,cuckercol}. Lagrangian
swarming models are essentially built upon rules extended from some or all of
the original Reynolds rules~\cite{reynolds}---Cohesion: moving towards the
average position of local flockmates; Alignment: steering towards the average
heading of local flockmates; Separation: avoiding crowding local flockmates.

Vicsek \textit{et al.}~\cite{vicsek} introduced a simple discrete-time model
of self-propelled particles with biologically motivated
interactions. Particles in that model move in a plane with constant speed
while aligning, at each time step, their velocity direction with their
neighbors' average direction of motion. Jadbabaie \textit{et al.}~\cite{jad}
provided the mathematical analysis and proof of convergence for Vicsek's
model. Couzin \textit{et al.}~\cite{couzin} developed a discrete model meant
to consider leadership and decision-making issues in animal groups. In
Couzin's model, at each time step, agents outside a given repulsion zone
follow the desired direction of travel by two acts: first by moving towards
the centroid of near neighbors, and second by getting aligned with the
velocity direction of agents in the local interaction range.
Olfati-Saber~\cite{rezaflock} introduced a flocking model based on a
behavioral algorithm embodying an extended form of the Reynolds
rules. Olfati-Saber's model is intrinsically continuous and has the
interesting and appealing ability of representing flock characteristics such
as rendezvous in space and obstacle avoidance. The Cucker--Smale flocking
model~\cite{cucker} assumes birds adjust their velocity through applying a
local linear consensus protocol which adds to the velocity a weighted average
of the differences of its velocity with those of the other birds. The entire
flock can therefore be represented by a complete weighted undirected graph
whose weights are a function of distance between every two individual birds or
nodes. The Cucker--Smale model can be either continuous or discrete. An
extension of that model that guarantees the collision avoidance property can
be found in Ref.~\cite{cuckercol}.

Another approach toward the study of collective behavior is based on an
analogy with the emergence of coherent behavior within a system of coupled
oscillators achieving synchronization. Watts and
Strogatz~\cite{watts98:_collec} studied the synchronization properties of
real-world networks, while Lago-Fern\'andez \textit{et
  al.}~\cite{lago-fernandez00:_fast} proved that clustering improves
synchronization. Small-world systems corresponding to identical oscillators
with linear coupling were studied by Barahona and Percora~\cite{barahona02},
while Nishikawa \textit{et al.}~\cite{nishikawa03} revealed that scale-free
networks are more difficult to synchronize compared to homogeneous networks. A
comprehensive application of this approach is given by Raley \textit{et
  al.}~\cite{naomi} with a particular focus on how a network of coupled
oscillators can be used to model the collective behavior of animals, with a
special emphasis on fish schooling. This continuous model supposes particles
can change their velocity heading but are unable to speed up or slow
down. More information on problems of synchronization involving complex
networks can be found in Ref.~\cite{arenas08:_synch}.

Despite these numerous efforts in developing continuous and discrete models,
very little insight has been gained into the structure and dynamics of the
information channel, which controls how information flows throughout the
swarm~\cite{sumpter}. Indeed, the vast majority of dynamical models reported
in the literature are primarily focused on devising refined behavioral
algorithms. The importance of deepening our understanding of this purely
decentralized architecture flow among system's components can be readily
acknowledged by recent discoveries of similar structures governing the very
mechanisms underlying social self-organization~\cite{camazine}.

In this paper, we bring together notions from ecology, network theory,
information theory, control theory, and agent-based modeling to establish and
comprehend the intricate relationship between the properties of the
information transfer channel---referred to as the swarm signaling network in
the sequel---and the dynamics of emergent collective behaviors based on local
interactions and decentralized control. Particular emphasis is placed on
gaining insight into: (i) what structurally makes swarming behaviors resilient
or robust, and (ii) how controllable the swarm can be. To this aim, we
explicitly define and construct the signaling network underpinning the group's
interactions that represents connections between all group members in the
physical space. This signaling network, channeling the flow of information
between agents, has a unique dynamics which is intimately connected to the
dynamics of the group members in the physical space. More specifically, we
show that the group's dynamic signaling network is composed of directed links
locally defined by a specific topological neighborhood of interactions for
each and every agent. The study of the connectedness of the swarm signaling
network allows us to uncover the pivotal relationship between swarm size and
number of neighbors in the topological neighborhood of interactions, which
proves to be in very good agreement with empirical observations obtained from
flocks of birds.  Using a dynamical model epitomizing our general framework,
we analyze swarming behaviors by thoroughly characterizing the dynamics and
structure of the signaling network. A profound connection between swarm
dynamics in the physical space and dynamics in the signaling network space is
uncovered. We find that swarm signaling networks are homogeneous and clustered
small-world networks---known to be prone to yielding large-scale
synchronization and emergence---even in the presence of environmental
noise. Subsequently, the resilience or robustness of the collective emergent
behavior is tested by adding exogenous noise in the environment. Depending on
the number of neighbors considered, using the $k$-nearest neighbor approach,
we show that consensus is achieved and maintained if the swarm signaling
network remains as a single giant strongly connected component at almost all
time. Finally, our analysis of the controllability of the swarm signaling
network enabled us to establish for the first time the analytical expression
of the number of driver nodes in terms of the swarm size and showing an
exponential decay with the number of nearest neighbors in the neighborhood of
interaction.

\section*{Results}

\subsection*{Connectedness of the signaling network}

%
Within our modeling framework (Methods section), the dynamic swarm signaling
network (SSN) is explicitly accessible and one may ponder over the details of
the relationship between connectedness of this network and emergent collective
behaviors through local synchronization. Here, we propose to bridge the gap
between two vastly different representations of the dynamics of our complex
adaptive system. On the one hand, we have the prevalent canonical
representation in the physical space---e.g. kinematic tracking of group
members---and, on the other hand, the SSN approach in the `network space'.

\begin{figure*}[htbp]
  \centering
  \includegraphics[width=0.95\textwidth]{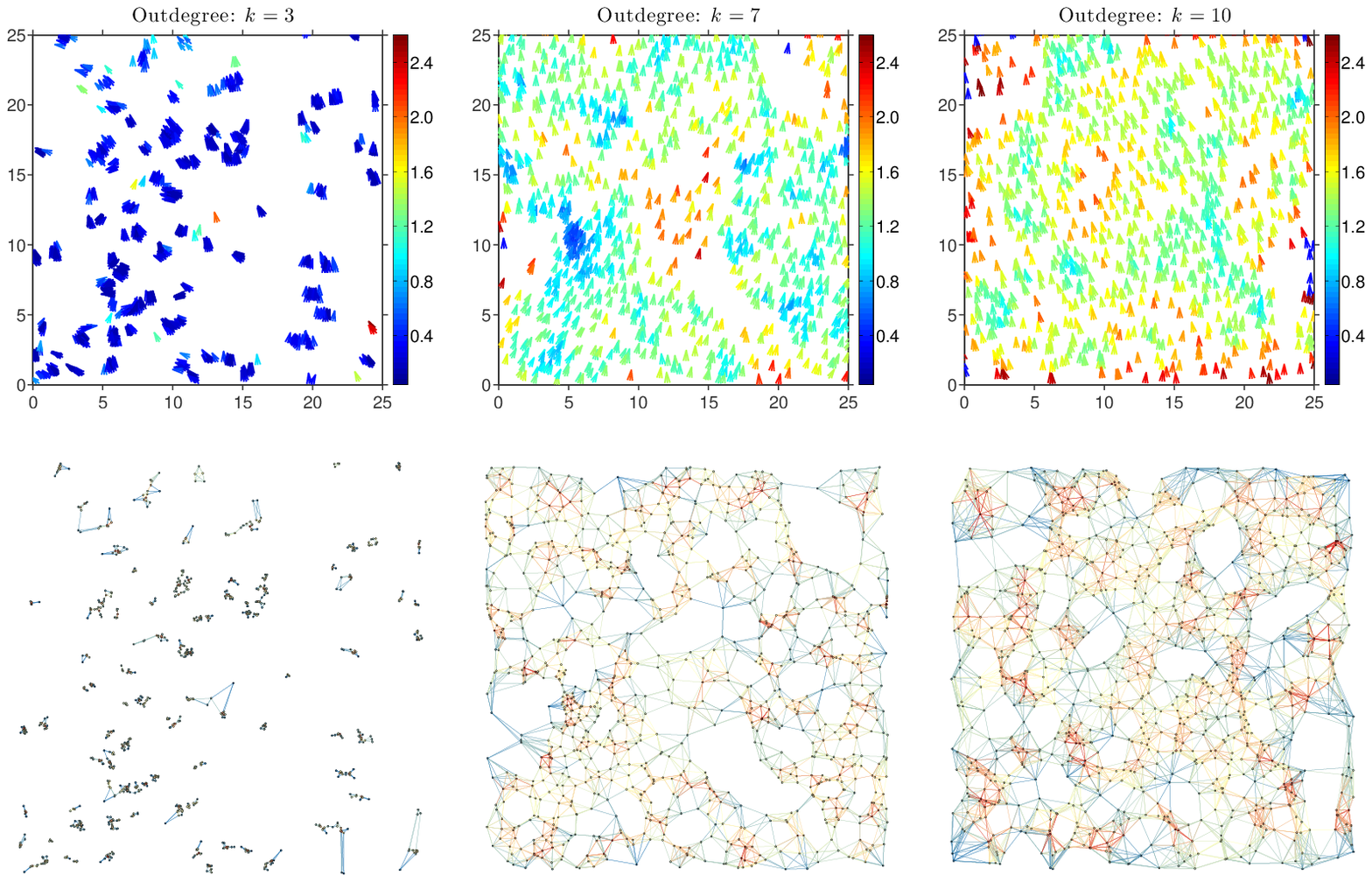}
  \caption{At a given instant, in a quasi-steady-state regime, velocity
    directions $\theta_i$ of $N=1000$ agents are displayed in the physical space (top row) and the
    associated SSN in the network space (bottom row) for
    three different values of the outdegree $k$: Left column: outdegree
    $k=3$; Center column: outdegree $k=7$; Right column: outdegree $k=10$. Top
    row: the actual velocity of an agent is indicated by a small arrow which color is
    mapped onto the size of the radius of the topological neighborhood of interactions. The
    vertical colormap is identical for all values of $k$, and the size of radius
    is expressed with the same spatial units as the square domain $[0,\
    25]^2$. Roughly, a blue arrow corresponds to an agent with a fairly small
    topological neighborhood of interactions, while, on the contrary, a red
    arrow indicates a large topological neighborhood of interactions.
    Bottom row: instantaneous SSN associated with the
    physical distribution of agents shown in the top row. The network nodes
    are exactly located at the agents' physical locations. The directed
    links are colored according to the value of the indegree $\kin$ of the
    source node, also colored, from which
    they are originating. A linear colormap ranging from blue to red is used
    with three different indegree intervals: $\kin\in [0,\ 8]$ for $\kout=3$,
    $\kin\in [1,\ 13]$ for $\kout=7$ and $\kin\in [3,\ 17]$ for $\kout=10$.
    The results correspond to the time step $t=3000$ nondimensional time units,
    which according to the results in Fig.~\ref{fig:GSCC}, is part of a quasi
    steady state. The noise level is fixed and set to
    $\eta_0=0.1\pi$ rad.
    \label{fig:noise}}
\end{figure*}
In the physical space, the emergent outcome appears before one's eyes
(Fig.~\ref{fig:noise} top row). Reaching local synchronization is a key factor
in forming a group and maintaining its emergent behavior, otherwise the group
will split apart unless a consensus is reached again. Furthermore, consensus
decisions bring along enhancement of decision accuracy compared with lone
individuals and improvement in decision
speed~\cite{ref:conradt,ref:quorum}. For a group to self-organize, the union
of the dynamically-evolving SSNs must have a spanning tree frequently
enough~\cite{ref:ren}.  Empirical evidences implicitly indicate the existence
of a signaling channel between every two arbitrary agents in the swarm at any
point in time. From the unique observations and findings of the STARFLAG
project, Cavagna \textit{et al.}~\cite{cavagna:scale} came up with this
compelling statement: ``The change in the behavioral state of one animal
affects and is affected by that of all other animals in the group, no matter
how large the group is''. Formally put, the SSN of the swarm is strongly
connected at all time which is a much stronger condition than the one
presented in Ref.~\cite{ref:ren}.

\begin{figure}[htbp]
  \centering
  \includegraphics[width=0.28\textwidth]{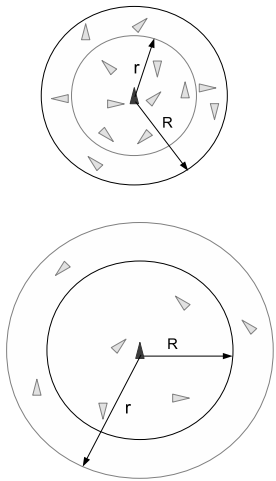}
  \caption{Schematics of metric (top) vs. topological (bottom) neighborhood of
    interactions. $R$ is the radius of the metric neighborhood and $r$ is the
    radius of the topological one based on the rule of $k$-nearest neighbors
    with $k = 7$. $R$ is constant as it defines a metric zone around the agent
    while $r$ changes in accordance with the distance between the agent and
    its $k$-th (here 7-th) nearest neighbor.  \label{fig:metric-topological}}
\end{figure}
\begin{figure}[htbp]
  \centering
  \includegraphics[width=0.48\textwidth]{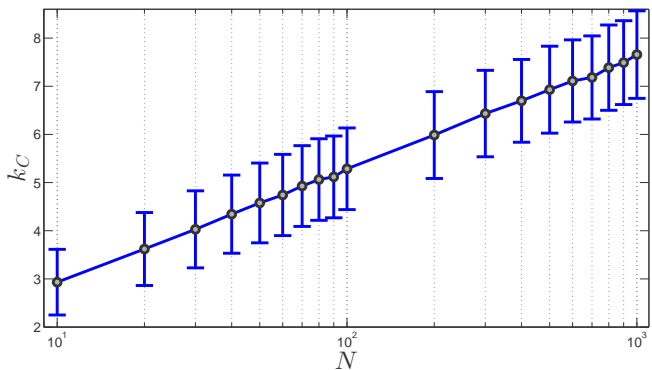}
  \caption{Critical value of the number of topological neighbors, $\kc$, for
    which the connectedness of the network is guaranteed, as a function of the
    swarm size $N$, with $N$ ranging from 10 to 1000. Grey dots represent the
    average value of $\kc$ obtained from a statistical analysis comprising
    1000 randomly generated $k$-nearest digraphs. The errorbars represent the
    associated standard deviations.}
  \label{fig:kC-N}
\end{figure}

The very first characterization of the SSN pertains to its connectedness,
which, in a $k$-nearest graph representing the topological interactions (see
Methods and Fig.~\ref{fig:metric-topological} for an introduction to the
differences between metric and topological neighborhoods), heavily depends on
the value of the outdegree $k$ (Fig.~\ref{fig:noise} bottom row). The
existence of a critical value, $\kc$, for the outdegree $k$ such that for
$k\geq \kc$ the $k$-nearest graph is connected, has never been
proved. However, Balister \textit{et al.}~\cite{ref:balister1} proved the
existence of $\kc$ in the probabilistic sense. More specifically, they proved
that for
\begin{equation}\label{eq:kc}
  k \geq \kc = c~\log N,
\end{equation}
where $N$ is the number of nodes---i.e. the number of agents in the
group---the probability for any randomly-generated $k$-nearest graph to be
connected tends to one. In Eq.~\eqref{eq:kc}, $c$ is a constant and the
smallest value found so far is $0.9967$~\cite{ref:balister1}. It is important
keeping in mind that those mathematical results were obtained under the
assumption that $N$ is large. When collective motion is considered, the number
of agents considered ranges from dozens to a few thousands, and rarely
more~\cite{viscek2012}. It is therefore important to assess numerically the
validity of Eq.~\eqref{eq:kc} for values of $N$ smaller than
1000. Figure~\ref{fig:kC-N} shows that even for small values of $N$, $\kc$
continues to scale linearly with $\log N$ on average. Moreover, the average
value of the coefficient $c$ here is found equal to $1.15$---this value tends
to decrease with increasing values of $N$, which is consistent with the value
$0.9967$ found in Ref.~\cite{ref:balister1} for large values of $N$.

Balister \textit{et al.}~\cite{ref:balisterrobust} further expanded this
result to the more conservative notion of $s$-connectivity. The SSN is said to
be $s$-connected if it contains at least $s+1$ agents, and the removal of any
$s-1$ of its agents does not disconnect it. Obviously, the concept of
$s$-connectivity is instrumental to study the resilience of our dynamic
SSN. Balister \textit{et al.}~\cite{ref:balisterrobust} found that for
$s\sim\log N$, the critical outdegree $\kc$ is asymptotically---i.e. for very
large swarms---the same for the $s$-connectivity as for the regular
connectivity. That is, as the outdegree $k$ is increased, the SSN becomes
$s$-connected very shortly after it becomes connected and the removal of a
small number of its agent will not harm the swarm's connectivity. This
property is consistent with a host of real-life observations on animal groups
in nature~\cite{ref:giardina,viscek2012}.

\subsection*{Structure of the signaling network}

\subsubsection*{Shortest connecting path}

\begin{figure}[htbp]
  \centering
  \includegraphics[width=0.48\textwidth]{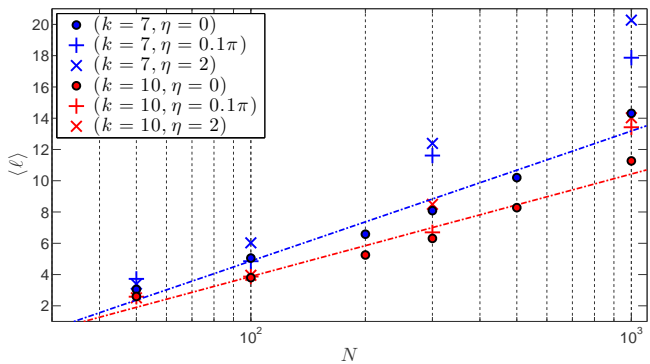}
  \caption{Average shortest connecting path vs. number of agents for the
    SSN. A log scale is used for the number of agents $N$. Two possible values
    of the outdegree are considered: $\kout=k=7$ and 10. Three values of the
    noise level $\eta$ are considered: noiseless ($\eta=0$), moderate
    ($\eta=0.1\pi$~rad), high ($\eta=2$~rad). The linear fitting in log scale
    is only shown for the noiseless case using dash-dotted lines.}
  \label{fig:shortest}
\end{figure}

%
Let us first consider the distance among agents in the swarm, and by distance
here we mean the network distance between nodes representing the agents in the
swarm network, and not the physical distance between agents in the physical
space. Typically this distance is defined by the shortest connecting path,
$\ell$, between any pair of agents. This metric is intimately related to the
small-world effect, with which it is possible to go from one agent to any
other in the swarm passing through a very small number of intermediate
agents. To be more precise, the small-world property refers to networks in
which the average shortest connecting path, $\langle \ell\rangle$, scales
logarithmically, or more slowly, with the number of agents $N$. Figure
\ref{fig:shortest} illustrates the average shortest connecting path $\langle
\ell \rangle$ versus $N$ for two different outdegree values $\kout=k=7$ and
$10$ for our SSN, and for three vastly different noise levels---noiseless,
moderate, and high. We chose those values for $k$ in order to ensure that the
network remains connected for up to $1000$ agents---the connectivity being
necessary to compute the average shortest connecting path. Given the log scale
on the $x$-axis, our results clearly confirm that the SSN exhibits the
small-world phenomenon for both values of the outdegree considered. Our
empirical result is further supported by a very recent mathematical analysis
by Alamgir \& von Luxburg~\cite{Alamgir}. Not surprisingly, a higher outdegree
shortens the shortest connecting path for all swarm sizes. On the contrary,
$\langle \ell \rangle$ is lengthened when the swarm evolves in increasingly
noisy environmental conditions, but the small-world property is conserved.

\begin{figure*}[htbp]
  \centering
  \includegraphics[width=0.95\textwidth]{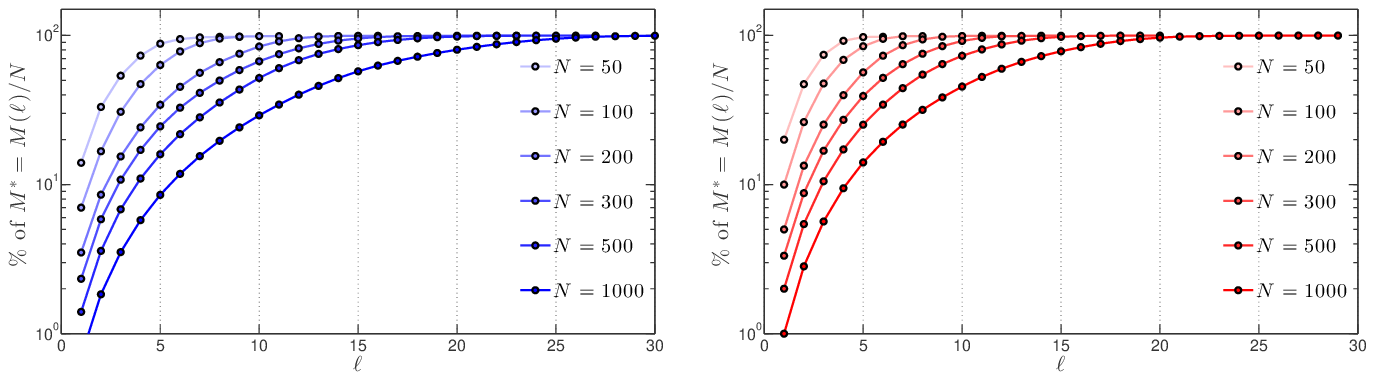}
  \caption{Normalized hop plot: $M^*=M(\ell)/N$ for the SSN. A log scale is
    used for the number of agents $M$ and various swarm sizes $N$ are
    considered. Two possible values of the outdegree are considered: Left:
    $\kout=k=7$; Right: $\kout=k=10$. The noise level is fixed and set to
    $\eta=0$.}
  \label{fig:M}
\end{figure*}

The small-world property can be more thoroughly analyzed by inspecting the
behavior of the quantity $M(\ell)$ defined as the average number of agents
within a network distance less than or equal to $\ell$ from any given
agent~\cite{ref:barrat}. The corresponding hop plot is shown in
Fig.~\ref{fig:M} for two values of the outdegree $\kout=7$ and $\kout=10$.
The exponential increase of $M$ with $\ell$ is yet another proof of the
small-world character of the SSN.

\subsubsection*{Clustering coefficient}

%
It is very interesting to observe that our swarm model (Methods section) based
on the $k$-nearest neighbor topological neighborhood of interactions (TNI:
Methods section and Fig.~\ref{fig:metric-topological}) generates a SSN
showcasing the small-world effect. However, in many social and technological
networks, the small-world effect is accompanied by a relatively high level of
clustering. For instance, random networks also exhibit the small-world effect
but possess an extremely low level of clustering.

\begin{figure}[htbp]
  \centering
  \includegraphics[width=0.48\textwidth]{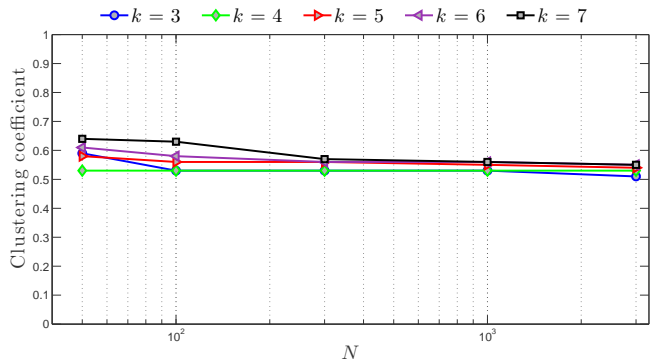}
  \caption{Clustering coefficient ($CC_{\textrm{\tiny out}}$) versus number of
    agents for the SSN. A log scale is used for the number of agents
    $N$. Different values of the outdegree are considered: $\kout=k=3, \cdots,
    7$. The noise level is fixed and set to $0$.}
  \label{fig:cc}
\end{figure}

The clustering coefficient, $CC$, characterizes the local cohesiveness of
networks~\cite{watts98:_collec} as well as the propensity to form clusters of
interconnected elements. Given the directed nature of the SSN and the fact
that neighbors are pointed at by outward edges, we consider the extended
definition of the clustering coefficient $CC_{\textrm{\tiny out}}$ given in
Ref.~\cite{ref:fagiolo}. Thus, the average clustering coefficient of our
$k$-nearest neighbor graph can be calculated as follows~\cite{ref:fagiolo}:
\begin{equation}
  CC_{\textrm{\tiny out}} = \frac{1}{k (k-1) N}~\textrm{trace}(A^2 A^\textrm{T}),
\end{equation}
where $k$, $N$, and $A$ are the outdegree, the number of agents, and the
adjacency matrix of the SSN, respectively~\cite{steen}. Figure~\ref{fig:cc}
shows the swarm's clustering coefficient as a function of the number of agents
$N$ in the swarm, for several different values of the outdegree $k$, and in
the absence of noise. These results highlight the rather high independence of
the clustering coefficient with both the number of agents and the
outdegree. We are therefore led to conclude that the SSN is intrinsically
highly clustered unlike random networks. Interestingly, those measured levels
of clustering are practically not affected by the presence of environmental
noise---moderate ($\eta=0.1\pi$~rad) and high ($\eta=2$~rad) noise levels were
tested. We contend that the high level of clustering in the SSN may find its
origins in the existence of clusters of agents in swarms, as commonly observed
in nature~\cite{krause02:_livin_in_group}.

\subsubsection*{Indegree distribution}

%
We have established that the SSN is a clustered small-world network. To better
understand its subtle structural organization, we now turn to the study of its
statistical homogeneity. Homogeneous networks are characterized by
fast-decaying degree distributions whereas heterogeneous networks produce long
and heavy tails---such power laws are a well-known signature of scale-free
networks~\cite{ref:barrat}.

\begin{figure}[htbp]
  \centering
  \includegraphics[width=0.48\textwidth]{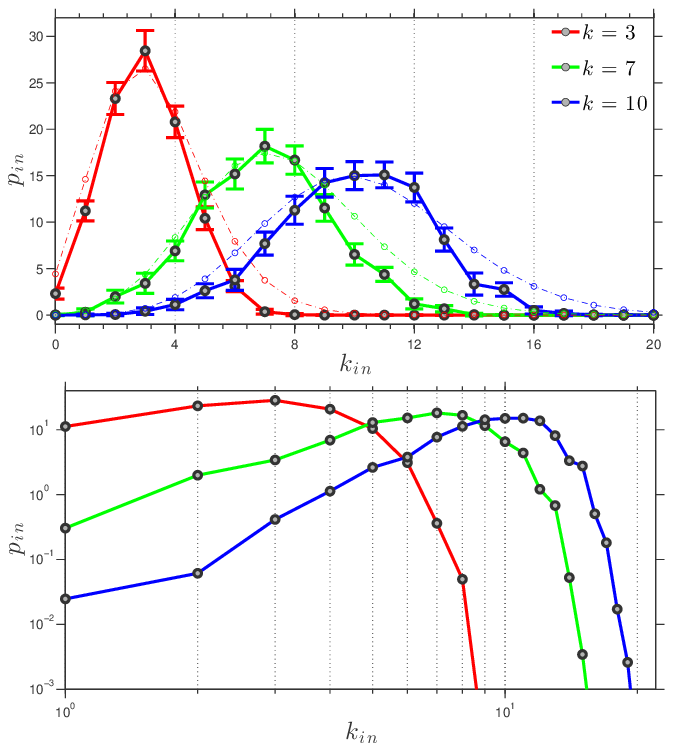}
  \caption{Indegree distribution $\pin$ of agents in the SSN for several
    simulations of the swarming model with $k=\kout = 3,\ 7,\ 10$ and
    different number of agents $N=50,\ 300$ and 1000; Top: linear scales with
    the exact values corresponding to the Poisson distributions for $k=3,7$
    and 10 shown using thin dash-dotted lines, and Bottom: logarithmic
    scales. The average indegrees $\langle \kin \rangle$ are $3, 7, 10$ and
    their standard deviations $\sigma_{k_{\textrm{\tiny in}}}$ are
    approximately $1.4,\ 2.2,\ 2.4$, for $k=\kout = 3,\ 7,\ 10$
    respectively. The noise level is fixed and set to $0$.}
  \label{fig:in--degree}
\end{figure}

The indegree, $\kin$, of an agent in the SSN is the number of directed edges
pointing at it; a directed edge representing a neighboring agent using the
information from the state of the agent that its edge is pointing at. The
indegree distribution, $\pin(\kin)$, is the fraction of agents in the SSN
having an indegree $\kin$. The average indegree distribution, $\langle \pin
\rangle$, for our SSN is computed for three distinct values of the outdegree,
$k=3,\ 7$ and 10. The averaging $\langle \cdot \rangle$ considered is a mixed
conditional averaging based on a temporal averaging of the network
configurations for 800 consecutive timesteps---with $\Delta t=1$---repeated 8
times each, and that for three different values of the total number of agents:
$N=50,\ 300$ and 1000. It is important to note that our results show very
little variation in the average indegree distributions for the three values of
$N$ considered. The results are shown in Fig.~\ref{fig:in--degree}, in which
the errorbars represent the standard deviation to the average value found.
The indegree distributions are peaked at $\kin=\kout=k$ for the three values
of the outdegree considered. More precisely, approximately half of the swarm
agents have an indegree such that $\kout -1 \leq\kin\leq \kout +
1$. Furthermore, for $k=7$ and $k=10$, the indegree distribution is
qualitatively symmetric about their maximum value obtained at
$\kin=\kout$. Based on the log-log plot of the indegree distribution in
Fig.~\ref{fig:in--degree}~(Bottom), it can be said that the indegree
distributions clearly are Poissonian like, with $\langle \kin \rangle =\kout$
and with a variance increasing with $\kout=k$. This is further verified by
comparing the results with the actual Poisson distribution as shown in
Fig.~\ref{fig:in--degree}~(Top) with a relatively good qualitative
agreement. Such Poissonian-like distributions are reminiscent of random
networks and starkly differ from power laws characteristic of scale-free
networks. Similarly to the clustering coefficient, measured indegree
distributions are practically not affected by the presence of environmental
noise---moderate ($\eta=0.1\pi$~rad) and high ($\eta=2$~rad) noise levels were
tested.

To further confirm the absence of an intrinsic characteristic scale for the
SSN, we computed the heterogeneity parameter $\kappa=\langle
\kin^2\rangle/\langle \kin \rangle$. Homogeneous networks are known to have a
$\kappa$ that scales with the indegree
$\kin$~\cite{ref:barrat}. Table~\ref{tab:kappa} shows the values of the
reduced heterogeneity parameter $\kappa^*=\kappa/\kin=\langle
\kin^2\rangle/\langle \kin \rangle^2$ for 9 SSNs corresponding to three values
of the outdegree $\kout=3,\ 7,$ and 10, and for 3 different sizes of swarms
corresponding to $N=50,\ 300,$ and 1000 agents. These results confirm the
homogeneity of all our SSNs as $\kappa$ indeed scales with the indegree
$\kin$, irrespective of the outdegree and swarm size. That allows us to
conclude that our SSNs are homogeneous and clustered small-world networks.
\begin{table}[htbp]
  \centering
  \begin{tabular}{l|ccc}
    $N$         & $\kout=3$ & $\kout=7$ & $\kout=10$\\
    \hline
    $50$   & 1.21 & 1.12 & 1.10\\
    $300$  & 1.21 & 1.10 & 1.06\\
    $1000$ & 1.31 & 1.09 & 1.08\\
    \hline
    \hline
  \end{tabular}
  \caption{Reduced heterogeneity parameter $\kappa^*=\kappa/\kin=\langle \kin^2\rangle/\langle \kin \rangle^2$ for 9 SSNs
    corresponding to 3 values of the outdegree $\kout=3,\ 7,$ and 10, and for 3
    different sizes of swarms corresponding to $N=50,\ 300,$ and 1000 agents.}\label{tab:kappa}
\end{table}

\subsection*{Resilience of the consensus}

\begin{figure}[htbp]
  \centering
  \includegraphics[width=0.48\textwidth]{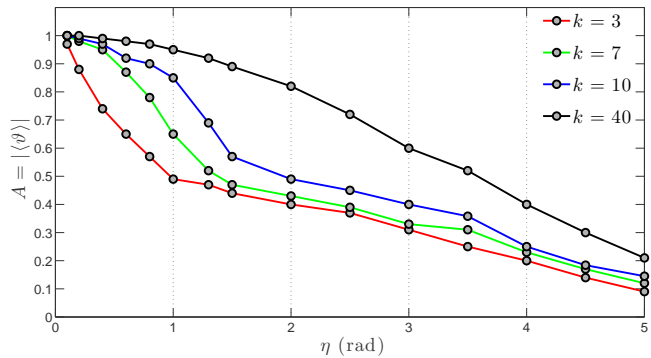}
  \caption{Alignment $A$ versus noise level $\eta$ for a swarm comprised of
    $N=1000$ agents. Three values of the outdegree are considered:
    $k=\kout=3,\ 7,$ and 10.}
  \label{fig:noiseeffect}
\end{figure}

%
The effects of noise on the dynamics of collective behaviors in the physical
space is well known and has been thoroughly investigated in the case of a
metric neighborhood~\cite{vicsek,viscek2012}. However, very little is known
about those effects in the case of a TNI, and more importantly on the dynamics
of the associated SSN. To this aim, we consider a swarm of $N=1000$ agents
evenly distributed throughout the physical domain, subjected to periodic
boundary conditions. Initially, all agents are heading North which globally
yields an alignment of unity. Figure~\ref{fig:noiseeffect} shows the impact of
noise on the alignment---i.e the consensus---of the swarm. In our framework,
the alignment is used as a measure of the resilience of the ordered phase of
the collective behavior to the effects of noise. As expected, the higher the
noise level $\eta$, the lower the alignment. For relatively low noise levels
$\eta$, the decay of the alignment is faster for lower values of the outdegree
$k$. For higher values of $\eta$, the decay of $A$ slows down and becomes
almost the same for the four values of the outdegree considered.

\begin{figure*}[htbp]
  \centering
  \includegraphics[width=0.95\textwidth]{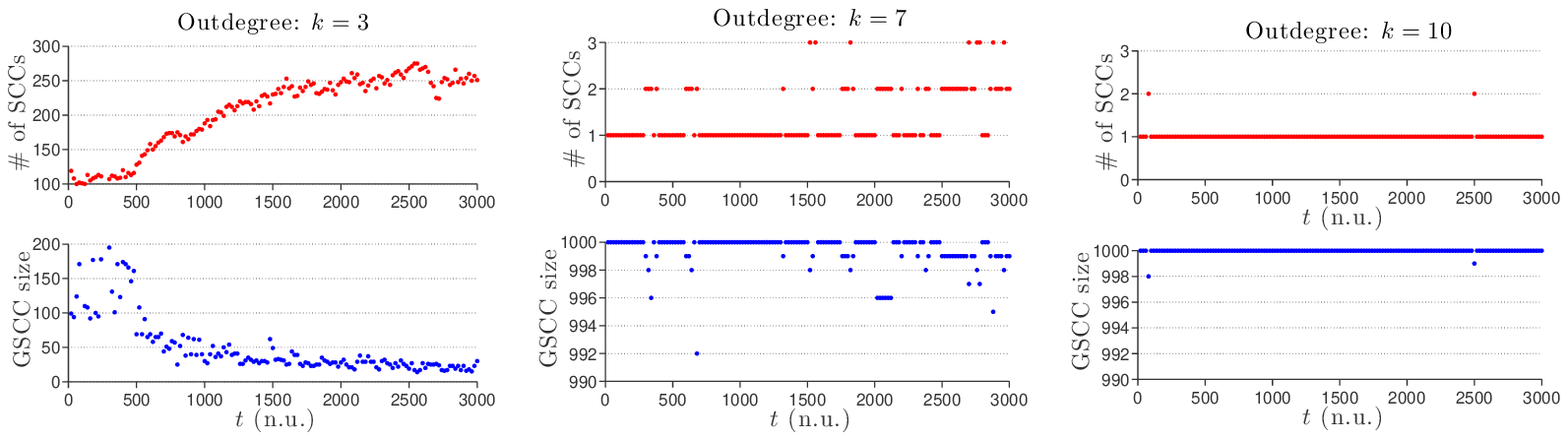}
  \caption{Dynamical properties of the GSCCs making the SSN. A dynamic range
    of 3000 nondimensional time units (n.u.) is considered with $N=1000$
    agents evenly distributed and all initially aligned with the North
    direction. The noise level is fixed and set to $\eta_0=0.1\pi$ rad. Top
    row: total number of SCCs. Bottom row: size of the GSCC found in the
    SSN. Left column: outdegree $k=3$; Center column: outdegree $k=7$; Right
    column: outdegree $k=10$.\label{fig:GSCC} }
\end{figure*}

The analysis of the SSN allows us to comprehend the above observations and
trends. We now fix the noise level at $\eta_0=0.1\pi$, which falls right into
the range where the alignment is significantly influenced by the outdegree. At
the very beginning, prior to any interaction, the SSN is strongly connected
for $k=7$ and $k=10$ and it forms a single giant strongly connected component
(GSCC) as is shown in Fig.~\ref{fig:GSCC} (top row). On the contrary, for
$k=3$ the SSN is composed of $114$ SCCs of very many different sizes: ranging
from $1$ agent to $99$ agents (Fig.~\ref{fig:GSCC}, top row). Another
informative quantity is the average neighborhood radius for the entire
swarm---the neighborhood radius is given by the largest distance separating a
given agent and its $k$ nearest neighbors. The initial average neighborhood
radii are $0.78$, $1.22$ and $1.49$ for $k$ equals to $3$, $7$ and $10$
respectively. We then let this complex system evolve through local
interactions of the agents and after a long-enough transient, the collection
of agents yields vastly different emergent behaviors in both the physical and
network spaces as shown in Fig.~\ref{fig:noise}.

For the low outdegree $k=3$, we observe a large number of clusters of
locally-aligned agents; no large-scale emergent coherent alignment is
achieved. This is clearly noticeable in both the physical and network spaces
(Fig.~\ref{fig:noise}, left column). The average TNI radius fell sharply from
$0.78$ to $0.21$ which is consistent with the physical
clustering. Furthermore, the dynamics has amplified the fragmentation of the
SSN, which, after the transient, contains 267 SCCs of much smaller sizes:
ranging from $1$ agent to $22$ agents (Fig.~\ref{fig:GSCC}, left column). Note
that the number of SCCs for $k=3$ tends to reach an asymptotic plateau about
the value 250 with very small-amplitude fluctuations after approximately 2000
nondimensional time units. We qualify this regime as quasi steady state.  On
the contrary, for both $k=7$ and $k=10$, a large-scale coherent alignment is
achieved while the distribution of agents is nonuniform but not as physically
clustered as in the case $k=3$. Those observations are corroborated by the
fact that the SSN remains as a single giant strongly connected
component---apart from very few agents splitting away from the ``peloton''
(Fig.~\ref{fig:GSCC}, center and right columns)---with almost unchanged
average TNI radii of $1.16$ and $1.44$ for $k=7$ and $k=10$
respectively. Furthermore, with a much larger value of the outdegree, $k=40$,
the swarm exhibits a higher level of resilience to noise with quite different
variations of the alignment with the noise level as compared to other smaller
values of $k$ considered.

\subsection*{Controllability of the signaling network}

%
If one wishes to control the dynamics of collective behaviors---a goal of
tremendous importance for both natural and artificial swarms, we now know that
it is necessary identifying the swarm's architecture, in other words the
SSN. From the engineering control viewpoint, such a dynamical system is said
to be controllable if it can be driven from any initial state to any desired
final state in finite time. Owing to the seminal work by Liu \textit{et
  al.}~\cite{ref:ctrl}, we know that it is first necessary to identify the set
of agents that, if driven by different signals, can offer full control over
the SSN. Liu \textit{et al.}~\cite{ref:ctrl} developed the analytical tools to
study the controllability of an arbitrary directed network allowing one to
identify the set of driver agents. Specifically, they proved that we can gain
full control over a directed network if and only if we directly control each
unmatched node---a node is said to be matched if a link in the maximum
matching points at it; otherwise it is unmatched--- and there are directed
paths from the input signals to all matched nodes.

The connectedness of the swarm signaling network is a sufficient condition for
an agent within the swarm to affect and get affected by some if not all agents
of the group. However, in many occasions, one or more agents need to be able
to drive the swarm to a certain global state, and usually within finite
time. This is better understood when considering two biological systems such
as a flock of birds or a school of fish. For instance, evasive maneuvers
triggered by predator or collision avoidance collective responses are induced
by one or a few agents perceiving the threat and responding to it. These few
agents effectively are driver agents in the abovedefined sense: they are able
to control the entire swarm by bringing the other agents to swiftly respond to
a threat that they are not directly detecting. It is worth adding that those
driver agents do not possess any ``super'' power of any sort but they simply
become drivers as they happened to have discerned the danger first; any other
agent in the swarm could be driving the group as long as it is subjected to
specific external cues which are not made available globally to the whole
swarm. In summary, for a specific dynamic collective behavior to occur,
connectedness and controllability of the SSN are necessary conditions.

A system's controllability is to a great extent encoded in the underlying
degree distribution, $p(\kin,\kout)$. That is, the number of driver agents is
determined mainly by the number of incoming and outgoing links each node of
the SSN has, and is independent of where those links point
at~\cite{ref:ctrl}. By construction the outdegree distribution of the SSN is a
Dirac delta distribution, while we found that its indegree distribution very
much resembles the one of a $k$-nearest random digraph. To allow for an
analytical study of the controllability of the SSN, we therefore consider the
following degree distributions:
\begin{align}\label{eq:dd}
  \pout(\kout)&=\delta(\kout-k),\\
  \pin(\kin)&= \frac{k^{\kin}}{\kin!}\text{e}^{-k}.
\end{align}
\noindent
\textit{Lemma.} The number of unmatched nodes of a graph having $N$ nodes and
a constant outdegree such that $\pout(\kout)=\delta(\kout-k)$, and an indegree
distribution of Poisson type $\pin(\kin)= \frac{k^{\kin}}{\kin!}\text{e}^{-k}$
is given by $\ND \approx \frac{N}{2} \text{e}^{-k}$, in the large $k$ limit.

\begin{figure}[htbp]
  \centering
  \includegraphics[width=0.48\textwidth]{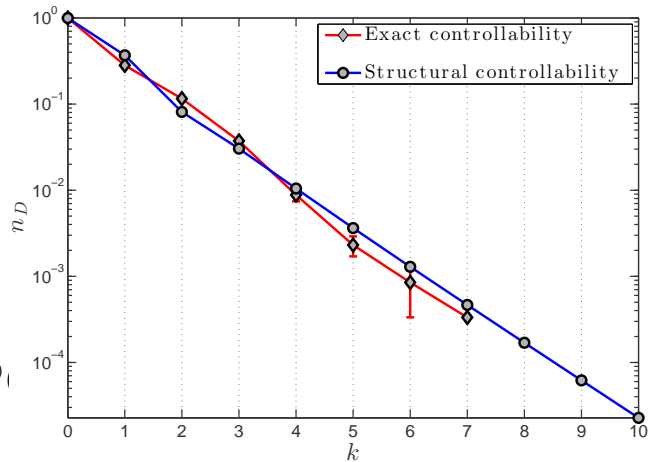}
  \caption{Density of driver agents, $\nD=\ND/N$, giving the proportion of
    agents necessary to control and drive a swarm of $N$ agents as a function
    of $k$, for a swarm dynamics with a topological neighborhood of
    interactions based on the $k$ nearest neighbors. The exact controllability
    framework is the one by Yuan~\textit{et al.}~\cite{yuan13:_exact}, while
    the structural controllability framework was developed by Liu \textit{et
      al.}~\cite{ref:ctrl}. Results using the exact controllability framework
    were obtained for 20 SSNs associated with a swarm of $N=3000$ agents for
    each data point;
    beyond $k=7$, $\nD$ drops to zero and the values are hence not shown. The
    average density of driver nodes was calculated and the associated standard
    deviations are shown using the errorbars. }
  \label{fig:nD}
\end{figure}

\begin{proof}
Following the approach developed by Liu \textit{et
  al.}~\cite{ref:ctrl}, the number of unmatched nodes, i.e. the minimum number
of driver nodes $\ND$ necessary to fully control the system, can be obtained
from the following generating functions
\begin{align}
  G(x) & = \sum_{\kout=0}^{\infty} \pout (\kout) x^{\kout}= x^k\\
  \hat{G}(x) & = \sum_{\kin=0}^{\infty} \pin (\kin) x^{\kin}= \text{e}^{-k(1-x)}\\
  H(x)  & = \sum_{\kout=0}^{\infty} Q (\kout+1) x^{\kout}= x^{k-1}\\
  \hat{H}(x) & = \sum_{\kin=0}^{\infty} \hat{Q} (\kin+1)
  x^{\kin}=\text{e}^{-k(1-x)},
\end{align}
where
\begin{align}
  Q(\kout)&=\frac{\kout \pout(\kout)}{\langle \kout \rangle}\\
  \hat{Q}(\kin) &=\frac{\kin \pin(\kin)}{\langle \kin \rangle}.
\end{align}
The general expression for the number of driver nodes $\ND$ obtained by Liu
\textit{et al.}~\cite{ref:ctrl} is given by
\begin{widetext}
\begin{equation}
  \nD=\frac{\ND}{N}=\frac{1}{2}\left\{
    \left[ G(\hat{w}_2) + G(1-\hat{w}_1)-1 \right] + \left[
      \hat{G}(w_2) + \hat{G}(1-w_1)-1\right]+k\left[ \hat{w}_1
      (1-w_2)+w_1(1-\hat{w}_2) \right] 
  \right\},\label{eq:nDgeneral}
\end{equation}
\end{widetext}
where, in the SSN framework
\begin{align}
  w_1&=H(\hat{w}_2)=\hat{w}_2^{k-1}\label{eq:1}\\
  w_2&=1-H(1-\hat{w}_1)=1-(1-\hat{w}_1)^{k-1}\label{eq:2}\\
  \hat{w}_1&=e^{-k(1-w_2)}\label{eq:3}\\
  \hat{w}_2&=1-e^{-kw_1}.\label{eq:4}
\end{align}
When $k=0$, the agents are totally independent and $G(x)=\hat{G}(x)=1$. Hence,
we trivially get $\nD=1$ from Eq.~\eqref{eq:nDgeneral}, which simply means
that we need to control 100~\% of the agents to control the dynamics of the
swarm---this conclusion is consistent with the noninteracting dynamics of the
group due to the choice of a 0-nearest neighborhood of interactions. We now
turn to the other pathological case, $k=1$, for which $w_1=1$, $w_2=0$,
$\hat{w}_1=\text{e}^{-1}$, $\hat{w}_2=1-\text{e}^{-1}$, such that
$\nD=\text{e}^{-1}\sim 0.368$.  For $k>1$, it is easy to check that
$w_1=\hat{w}_2=0$ are the smallest roots for $w_1$ and $\hat{w}_2$ in the
system of Eq.~\eqref{eq:1} and Eq.~\eqref{eq:4}. Hence, the fraction of driver
nodes simplifies to
\begin{align}
  \nD&
  =\frac{1}{2}\left[G(1-\hat{w}_1)-1+\hat{G}(w_2)+k\hat{w}_1(1-w_2)\right],\\
  \intertext{or more explicitly} \nD&=
  \frac{1}{2}\left[\left(1-\text{e}^{-k(1-w_2)}\right)^k
    -1+\text{e}^{-k(1-w_2)} +k\text{e}^{-k(1-w_2)}(1-w_2) \right],\label{eq:nD0}\\
  \intertext{in which $w_2$ is solution of the self-consistent equation}
  1-w_2&=\left(1-\text{e}^{-k(1-w_2)}\right)^{k-1}.
\end{align}
With those results, $\nD$ can easily be calculated and results are shown in
Fig.~\ref{fig:nD}. The asymptotic behavior of $\nD$ in the large $k$ limit can
easily be determined as $w_2$ tends to 1. Hence, at the leading order
\begin{equation}\label{eq:nd}
  \nD \approx \frac{1}{2} \text{e}^{-k},
\end{equation}
which appears very clearly on the graph in Fig.~\ref{fig:nD} given the log
scale on the $y$-axis. This concludes the proof of the above Lemma.
\end{proof}

It is important noting that within the structural controllability framework
developed by Liu \textit{et al.}~\cite{ref:ctrl}, binary link weights such as
those considered in the SSN (see Methods section and Eq.~\eqref{eq:consensus})
cannot be considered per se as they must be free independent parameters. This
issue can readily be resolved by considering the more realistic case of
non-binary weights accounting for the imperfections of the information
transfer channels through which the agents interact. Alternatively, one may
consider the exact controllability framework very recently developed by
Yuan~\textit{et al.}~\cite{yuan13:_exact}, which offers a more universal tool
to evaluate the controllability of any complex network. As is shown in
Fig.~\ref{fig:nD}, the results from both frameworks---structural
controllability and exact controllability---are fully consistent.

The last question that should be answered regarding the above result on the
number of driver nodes and the overall controllability of the SSN lies with
the dynamic nature of the SSN. Since the SSN is intrinsically a switching
network---at each instant a certain number of links are broken while the exact
same number of edges are created due to the motion of the agents in the
physical space---one can prove using Eq.~\eqref{eq:nd} that it is controllable
at each instant, assuming of course a high-enough value of $k$. If that is the
case, it is known from control theory associated with dynamic multi-agent
systems that the overall switching dynamical system is
controllable~\cite{switchctrl,switchctrl2}.

\section*{Discussion}

\begin{figure}[htbp]
  \centering
  \includegraphics[width=0.48\textwidth]{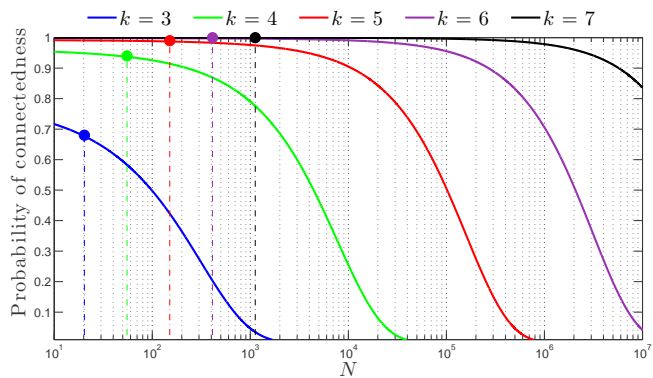}
  \caption{Probability of connectedness for the SSN vs. number of agents $N$
    for different values of the number of nearest neighbors $k$. The SSN
    corresponds to a specific configuration of the swarm in which $N$ nodes
    are placed in a unit square independently through a uniform
    distribution. Then each node is connected to its $k$ nearest neighbors to
    form the $k$-nearest graph. For each value of the outdegree $k$, the
    maximum size of the swarm population $\Nc$---given by $k=\kc=c\log \Nc$
    with $c=0.9967$~\cite{ref:balister1}--- ensuring the connectedness of the
    SSN is represented by a colored dot with the associated vertical dashed
    line.}
  \label{fig:balister}
\end{figure}

%
The study of the connectedness of the SSN allowed us to uncover the existence
of a relationship between the swarm size, given the number $N$ of agents, and
the number $k$ of nearest neighbors influencing any agent's behavior and
dynamics. Indeed, the general results from graph theory applied to the study
of the SSN connectedness take a particular significance in the context of
dynamic collective behavior where the number of agents $N$ may not necessarily
be very large and the number of nearest neighbors, $k$, cannot possibly exceed
at most 15 to 20 due to the intrinsic bandwidth limitations in signaling,
sensing and internal information processing. To better appreciate these
results, we present in Fig.~\ref{fig:balister} the dependence of the
probability of connectedness of the SSN as a function of $N$ for different
values of $k$. Despite the uniform character of the distribution of agents in
the swarm considered to establish Fig.~\ref{fig:balister}, this figure reveals
the profound relationship between connectedness of the swarm and the number of
agents $N$, for different values of the outdegree $k$. This result was already
suggested by Eq.~\eqref{eq:kc}. For the sake of explanation, let us consider a
swarm comprised of $N=1000$ agents, which is a reasonable number for living
animals~\cite{krause02:_livin_in_group}. Figure~\ref{fig:balister} shows that
this swarm will remain connected at all time if $k$ has at least a value of
approximately 6 or 7. This result is in very good agreement with the
experimental observations of Ballerini \textit{et al.}~\cite{cavagna} for
flocks of starlings with approximately $1000\sim 1200$ birds at maximum. Based
on their thorough analysis of the dynamics of flocks, Ballerini \textit{et
  al.}~\cite{cavagna} claimed that each starling had a TNI made up of 6 to 7
other birds. Thus, our model leads to a more general rule of interaction in
swarms: each agent interacts on average with a fixed number of neighbors
irrespective of the distance, and that number of neighbors $k$ depends on the
swarm size $N$. By extension, for artificial swarms, which typically have a
much smaller size---with say $N$ being at most 100---our analysis enables us
to conclude that 4 to 5 interacting neighbors are necessary to ensure the
swarm's connectedness and effectiveness. Note that, this analysis based on
Fig.~\ref{fig:balister} does not account for the dynamics of the SSN and more
importantly for the ubiquitous presence of noise in the environment.

Beyond the connectedness of the SSN, we found for the first time the details
of its structural properties revealing that, if connected, the SSN is a
homogeneous and clustered small-world network even when considering the
disruptive effects of noise on the inter-agent interactions. Hence, the swarm
information transfer channel has a relatively high local cohesiveness and no
intrinsic characteristic scale could be found in the indegree
distribution. The small-world phenomenon could have been intuited through the
mere observation of exceptionally fast responses of biological swarms to
external cues, e.g. fish school evasive maneuver, collision avoidance,
etc. The homogeneous character of the SSN could also have been
intuited. Indeed, the difference in indegree distribution has vastly
significant implications for the structure of the networks. For instance, the
long tail of power-law distributions of the indegree is a clear signature of
the existence of hubs in scale-free networks. Interestingly, even though our
swarm network is not, per se, a random network---its dynamics is governed by a
set of rules, including the $k$-nearest neighbor rule---its indegree
distribution is not able to reflect those differences with real random
networks. Note that, this result is not surprising given that we are dealing
with a collection of identical agents with a very minimal level of state
properties; a power-law signature with the associated hub effect seems
unthinkable in our context. However, we nonetheless observe that some specific
agents do ``attract'' much more attention than others with indegrees of 15 and
above (Fig~\ref{fig:in--degree}). Finally, it is interesting comparing the
structural properties of the SSN based on a TNI with the ones for a signaling
network based on a metric distance. Both interaction distances lead to similar
levels of clustering and similar average shortest connecting paths. The
central difference between the two groups of SSNs lies with the fact the
topological SSN is a directed network while the metric SSN is undirected. As a
direct consequence of that, the outdegree distributions of both types of SSNs
are fundamentally different: the outdegree of the topological SSN is constant
and equal to $k$, while the outdegree of the metric SSN is identical to the
indegree distribution, which we found to be Poissonian-like.

A central point to always keep in mind is the fact that the SSN has a dynamics
that is evolving hand in hand with the dynamics of the agents
themselves. Hence, the connectedness and the structural properties of the SSN
are in general not constant. Our analysis reveals this profound connection
between, on the one hand, the dynamics of the collection of agents in the
physical space and the structural properties of the SSN as well as its own
dynamics, on the other hand. This comment is very elegantly epitomized by
Fig.~\ref{fig:noise} which stresses the parallel between the structure of the
swarm in the physical space and the associated SSNs for the three different
values of the indegree considered, namely $k=3,\ 7$, and 10. The instantaneous
SSNs associated with the physical distribution of agents are shown in
Fig.~\ref{fig:noise}, bottom row. The network nodes are exactly located at the
agents' physical locations, and the directed links are colored according to
the value of the indegree, $\kin$, of the source node from which they are
originating. For instance, we are able to visually correlate high values of
the indegree $\kin$ to small radii of the TNI. A better understanding of this
observation would of course require a more thorough analysis which is beyond
the scope of the present study. Another point has to be made about the
connection between SSN structure and swarm dynamics in terms of consensus
speed. Intuitively, one can easily imagine that a larger number of topological
numbers $k$ leads to faster consensus since the connectivity of the network
underpinning the dynamics of the interacting swarming agents affects
profoundly the consensus capability---in general, higher degree of
connectivity yields higher rate of convergence to consensus
\cite{reza,wenwuyu,yasaminmos,aragues12}. This fact has very recently been
proved exactly by Shang \& Bouffanais~\cite{shang13:_influen}. However it is
important to note that adding more edges by increasing the number of
topological agents with whom one is interacting is feasible but only up to a
certain extent as there is always a cost associated with information exchange
and also due to inherent limits in terms of signaling mechanisms, sensory and
cognitive capabilities---for instance, see
Ref.~\cite{emmerton93:_vision_brain_behav_birds} for such biological
considerations with pigeons and Ref.~\cite{tian09:_optim} for SPPs having a
limited view angle. 

In our framework we considered the simplest
topological model of all consisting in having the same number of nearest
neighbors $k$ for all agents. Obviously, this framework can be extended in
many ways but one particular extension is worth mentioning: the case where $k$
varies from agent to agent depending on some local parameters, e.g. the
neighbors density of neighbors, the size of TNI radius, etc. Such a local
adaptation of the value of the outdegree $k$ clearly enforces a very specific
outdegree distribution. Some very recent works on the controllability of
complex networks~\cite{jia1,jia2} allow to conclude that this would have a
direct impact on the swarm controllability. Hence, this leads to the following
intricate inverse problem of finding one or more distributions of $k$ generating an
optimal controllability of the swarm.

From the practical standpoint of designing artificial swarms, our knowledge of
the properties and dynamics of the SSN, and their influence on the swarm
dynamics is necessary but not sufficient. Gaining a better understanding of
its controllability is paramount. Through Eq.~\eqref{eq:nD0} and
Eq.~\eqref{eq:nd}, we have analytically established that the number of driver
nodes decreases exponentially as the number of nearest neighbors
increases. Note that for a metric-based SSN, the density of driver nodes is
easily obtained as $\nD \approx \textrm{e}^{-\langle k
  \rangle/2}$~\cite{ref:ctrl}. In addition, the value $r$ of the radius
defining the metric neighborhood conditions the value of the mean degree
$\langle k \rangle$. If one chooses a topological neighborhood such that
$k^{\textrm{\scriptsize T}}=\langle k^{\textrm{\scriptsize M}} \rangle
$---where the superscript ``T'' refers to topological and ``M'' to
metric---then the topological SSN can be said to be more controllable as $\nD$
decreases faster with $k$ as compared to the metric case. Note that in the
case of hierarchical group dynamics such as those reported by Nagy \textit{et
  al.}~\cite{nagy}, the signaling network has a well-defined tree
structure. The controllability of such networks has been analytically
established in Refs.~\cite{zamani,bo}.

We can say that if the number of nearest neighbors reaches a value of $6$ or
$7$---for instance considering a flock of birds like those studied in the
field by Ballerini \textit{et al.}~\cite{cavagna}---every agent not only
affects and is affected by all other agents within the group, but more
importantly, is capable of full control over all other agents. More generally,
when a large swarm is considered its effectiveness and resilience entail the
connectedness of the SSN. From Eq.~\eqref{eq:kc}, we can consider that the
number of interacting neighbors is at minimum $\kc=c~\log N$, hence leading to
$\nD \sim 1/N^c\ll 1$ using Eq.~\eqref{eq:nd}. This result proves that
ensuring the connectedness of large swarms automatically ensures its full
controllability. However, it is possible that this interesting result ceases
to be true for very small swarms.  In summary, this ability to control the
swarm is instrumental in situations where an agent---or even a few number of
them---needs to play a leadership role in guiding the swarm either toward a
certain destination or away from a potential danger. Note that this leadership
role can be temporary or permanent.

\section*{Methods}\label{sec:Methods}

\subsection*{General features of the model}

%
Here, swarming refers to a circumstance in which multiple adaptive agents---be
them living creatures or artificial ones---create a certain level of
spatiotemporal order characterized by one or more macro-level properties. For
the sake of clarity, we consider a collective of $N$ locally-interacting
adaptive and identical individuals. Each individual agent $i$, at any given
instant $t$, is assumed to be fully characterized by the state variable
$\psi_i(t)$. Such a generic state variable may represent widely different
characteristics depending on the nature of the group considered: e.g. employed
or unemployed forager state for honey bees, kinematic variables for fish in a
school, birds in a flock or robots in an artificial swarm, space available for
a pedestrian on a congested sidewalk, etc.

The nonlinear dynamics of each agent $i$ takes the general form
\begin{equation}\label{eq:state}
  \frac{\textrm{d}\psi_i(t)}{\textrm{d}t} =
  f(\psi_{j}(t),\psi_{j+1}(t),\ldots,\psi_{j+k-1}(t),\psi_i(t)),
\end{equation}
that stresses the local nature of the interactions between agents since the
subset $\Psi_i(t)=\{ \psi_m\}_{m=j,\cdots,j+k-1}$ only includes a fraction $k$
of the $N$ agents affecting the behavior of agent $i$. Note that the formalism
of Eq.~\eqref{eq:state} does not capture the fact that the value of the $k$
indices---from $j$ to $j+k-1$ above---are actually $i$-dependent since they
are defined by the belonging, or not, of an agent to the neighborhood of
interaction of agent $i$. Moreover, these $k$ indices may change over time due
to the dynamical nature of the neighborhood of interactions, itself imposed by
the dynamics of agent $i$. That means that in general, the makeup of $\Psi_i$
varies from individual to individual and changes with time. Specifically, it
is entirely dependent on how the neighborhood of interactions---formally
represented by $\Psi_i$---is constructed which further defines the
communication links between agents. The neighborhood of interactions is the
cornerstone of the global SSN, and its intricate structural properties and
dynamics have been studied below. Moreover, the values of each $\psi_m$ within
$\Psi_i$ are made available to the internal control processing mechanism
through the various sensory modalities defining multiple communication
channels between group members---e.g. mechanical signaling through lateral
line sensing and visual signaling are both involved in fish
schooling~\cite{krause02:_livin_in_group}. The function $f$ in
Eq.~\eqref{eq:state} embodies the specifics of each individual's internal
control processing mechanism. It is worth highlighting at this stage that
complex collective dynamics can be achieved with simple $f$ given the possibly
nontrivial dynamics of $\Psi_i$ depending on the very nature of the
neighborhood of interactions.

At this point, we make another general assumption consisting in imposing that
any decision made by a group member is based on relative state values and not
on absolute ones. If the state variable $\psi_i$ is a quantity that is frame
dependent, such as the agent's velocity, the agent is solely able to
appreciate an interacting neighbor's state with respect to its own. This
argument may even hold for non-frame dependent state
variables---e.g. pheromone levels in ant trails---and is easily reconcilable
with the multiple gradient-based taxes observed in many biological
systems~\cite{dusenbery92:_sensor_ecolog}. Formally, this relative-state
assumption reads
\begin{equation}\label{eq:statediff}
  \frac{\textrm{d}\psi_i(t)}{\textrm{d}t} =
  g(\psi_{j}(t)-\psi_i(t),\ldots,\psi_{j+k-1}(t)- \psi_i(t)).
\end{equation}
The function $g$ is referred to as a consensus protocol---intrinsically local
by the nature of its inputs $\tilde{\Psi}_i(t)=\{
\psi_m-\psi_i\}_{m=j,\cdots,j+k-1}$---if a steady-state can be reached and
once it is reached, if the following relations hold: there exists a function
$h$ such that
\begin{equation}
  \psi_i(t) = \cdots = \psi_N(t) = \textit{h}(\psi_i(0), \cdots ,\psi_N(0)),
\end{equation}
where $\psi_i(0), \ldots ,\psi_N(0)$ are agents' initial state conditions,
e.g. agents' initial velocity directions in Ref.~\cite{reza}. In simple words,
the local synchronization protocol defines for each individual agent what
Sumpter~\cite{sumpter} calls the behavioral algorithm, also known as the
internal information processing mechanism responsible for the behavioral's
response to the sensed external information that is flowing in a decentralized
way throughout the swarm.

\subsection*{Topological neighborhood of interactions}\label{sec:topological}

%
We now aim at formalizing the key concept of neighborhood of
interactions. From our introduction above, it appears clearly that
$\tilde{\Psi}_i$ fundamentally depends on a series of factors which include:
signaling mechanisms, sensory and cognitive capabilities. The signaling
mechanisms are the different vehicles for the information to flow through the
swarm's surrounding environment. The sensory capabilities are responsible for
information acquisition from the surrounding environment to the internal agent
domain. Within that domain, the internal information processing is taken care
of by the cognitive capabilities. Even though the information chain has been
clearly identified, we believe that accurately modeling each and every
component is nonessential. Indeed, one and only one of those components will
be the limiting factor and depending on the environmental conditions, that
limiting factor may change; e.g. fish schooling from crystal-clear waters to
murky ones~\cite{cheng11:_aggreg_patter_trans_sligh_varyin}. Therefore, we
consider a topological neighborhood of interactions (TNI)~\cite{lukeman} whose
physical relevance was discussed in
Ref.~\cite{ginelli10:_relev_metric_inter_flock_phenom}.

The vast majority of models of collective animal behaviors found in the
literature are based upon a metric neighborhood of interactions. In that
specific class of models, the only thing that matters for an agent is the
physical distance to neighboring agents. A typical example of an agent's
metric neighborhood is the open ball interaction zone with radius $R$ centered
about the agent. The simplicity of the metric-based neighborhood approach is
evident and that translates into a relative ease of computational
implementation. However, it suffers from many limitations; for instance it
cannot account for the cognitive limitations of agents evolving in very dense
crowds~\cite{krause02:_livin_in_group}.

European project named Starlings in Flight or STARFLAG has been one of the
most recent and largest experiments in the human history carried out to
analyze the collective behavior of birds~\cite{cavagna}. By reconstructing the
three-dimensional positions of individual birds in airborne flocks of a few
thousand members, Ballerini \textit{et al.} show that the interaction does not
depend on the metric distance, as most current models and theories assume, but
rather on the topological distance. They discovered that each bird interacts
on average with a fixed number of neighbors (six to seven), rather than with
all neighbors within a fixed metric distance. To the best of our knowledge, an
explanation for this surprising empirical observation has yet to be
given. Ballerini \textit{et al.}~\cite{cavagna} claim that interactions based
on metric distance is unable to reproduce the density changes, typical of bird
aggregations, because one would expect cohesion to be lost when mutual
distances become too large compared with the interaction range. In addition,
with social networks, the relevance of the topological distance between
neighbors becomes apparent and it is believed that it could determine how
populations move in, split up and form separate groups~\cite{bode,borrel}. For
instance, guppies preferentially shoal with individuals of a similar
size~\cite{croft}, and faster individuals are more likely to be found at the
front of groups~\cite{wood}.

With a TNI, one has to be watchful for the possibility of the topological
distance becoming too large so that the interaction or information exchange
could not take place. In practice, that can potentially happen with very low
density swarms or when some individual agents become widely separated from the
swarm. In our numerical framework, the existence of periodic boundary
conditions combined with a relatively high density of agents prevent such
extreme case from happening. Still with a TNI, an agent is not just concerned
about the physical distance to its neighbors. Many other diverse and subtle
aspects can be factored in, such as the maximum number of neighbors set by
some cognitive limitations, familiarity and other social relationships,
etc. The rule of $k$--nearest neighbors~\cite{parrish} epitomizes the
topological paradigm. Figure \ref{fig:metric-topological} illustrates and
highlights graphically some of the fundamental differences between a metric-
and a topological-based neighborhood of interactions---the rule of
$k$--nearest neighbors is considered. The metric neighborhood or interaction
zone is an open ball with a constant radius, $R$, centered about the agent
while $r$, the radius of the TNI, has an adaptive behavior to include the
$k$-th (here 7-th) nearest neighbor. It is apparent that $r$ is not just a
function of the physical distance.

\subsection*{Swarm signaling network}

%
Let us consider members of a swarm, say a few hundreds, heading towards a
certain destination. An individual agent lagging behind the large swarm,
isolated from those moving together, decides to join the mainstream. Some
information from the agents in the bulk of the swarm will flow towards the
lonely agent and will almost surely affect its migratory behavior. Whereas
agents within the swarm will most probably receive no information from the
loner and will therefore experience no change in their behaviors. This
phenomenon simply reflects the directed nature of the interactions among
agents. Apart from this revealing case, empirical evidences support the idea
of directed interactions in pigeon flocks~\cite{nagy}.

We now precisely define and construct the SSN which, as already mentioned, is
the information transfer channel underpinning the dynamics of the interacting
swarming agents. Constituent links of the SSN of a group whose agents have
directed interactions are unidirectional by opposition to bidirectional
interactions in a group of agents with undirected interaction edges. The TNI
based on the $k$-nearest neighbor rule allows one to locally identify the
links between agents. The topological character of the neighborhood of
interactions has a tremendous impact on the properties of interagent
connectivity, in particular with the induced asymmetry in the relationship
whereby if agent $j$ is in the neighborhood of agent $i$, then $i$ is not
necessarily in the neighborhood of $j$, i.e. the interaction is directed. On
the contrary, with a metric neighborhood the interagent connectivity is
fundamentally symmetric with the presence of undirected interactions.

Through a bottom-up assembly of the interagent links, the complete global
graph characterizing the connectivity can be constructed. Given the dynamics
of the TNI and the directed nature of the links, the SSN is a switching
strongly connected $k$--nearest neighbor
digraph~\cite{ref:eppstein,ref:balister1,ref:balister2}. It is worth noting
that the random graph
theory~\cite{ref:barrat,ref:handbook,ref:scc,ref:robust,ref:core} is not
appropriate, nor relevant to the study of the dynamics of the connectivity in
swarms since links are introduced irrespective of any distance between
nodes---be that in the physical space or in the signaling network space.

\subsection*{Dynamic swarming model}

%
Above, we emphasized the generality of the concepts at the core of our
modeling framework. Thus, details such as the nature of the state variables or
the type of interactions between agents were intentionally left out. We
believe that those specific details do not have an impact on the key features
at the heart of emergence in collective behaviors; this approach can be
regarded as a ``crude look at the whole'' as advocated by the Physics Nobel
Laureate Murray Gell-Mann~\cite{gell-mann96}.

To exemplify our general framework for collective behaviors, we consider
self-propelling agents moving about a two-dimensional plane with constant
speed, $v_0$, similarly to Vicsek's model~\cite{vicsek}. However, our
neighborhood of interactions is not metric but instead is topological. For
simplicity, we assume that each agent $i$ is fully characterized by one unique
state variable $\psi_i$, its velocity $\mathbf{v}_{i} = v_0 \cos\theta_i
\hat{x} + v_0 \sin\theta_i \hat{y}$, or equivalently its velocity direction
$\theta_i$, the speed $v_0$ being constant. The local synchronization
protocol, based on relative states and generically stated as in
Eq.~\eqref{eq:statediff}, is strictly equivalent to a local alignment rule
which mathematically can be stated as:
\begin{equation}\label{eq:consensus}
  \dot{\theta_i}(t) = \frac{1}{\left| \mathcal{N}_i(t) \right|}\sum\limits_{j \in
    \mathcal{N}_i(t)} w_{ij}(\theta_j(t) - \theta_i(t)),
\end{equation}
where $\mathcal{N}_i(t)$ is the time-dependent set of outdegree neighbors in
the TNI of agent $i$, with cardinal number $\left| \mathcal{N}_i(t) \right|$,
and $w_{ij}$ is the binary weight of the $i-j$ communication link. Note that
in some models, $w_{ij}$ can take a more complicated form than our binary
choice~\cite{mirabet,cucker,bode2}. Using the $k$-nearest neighbor rule for
the TNI, we have $\left| \mathcal{N}_i(t) \right|=k$ and the following
dynamical equation for each individual agent in the swarm:
\begin{equation}\label{smotion}
  \dot{\theta}_{i} = \frac{1}{k}\left[(\theta_j - \theta_{i}) + \cdots +
    (\theta_{j+k-1}  - \theta_{i})\right],
\end{equation}
where $\theta_j,\cdots,\theta_{j+k-1}$ are its $k$-nearest neighbors' velocity
directions. The dynamics of the agents in the two-dimensional physical space
are intricately coupled to the dynamics of the SSN. This network is, by
construction, a switching $k$-nearest neighbor digraph, for which the specific
value of $k$ has a direct impact on its strongly connected character.

Up to this point, our modeling framework is based on a continuous-time
approach. From a practical standpoint, it is necessary switching to a
discrete-time approach; the associated sampling time, $\Delta t$, being
intimately connected to some of the characteristic physical times of our
complex dynamical system: e.g. agent's speed, speed of interagent information
exchange, speed of internal information processing within one agent, etc. Once
a sampling time $\Delta t$ has been selected or is imposed by the natural or
artificial characteristics of the system, the set of equations governing the
discrete-time dynamics of the agents' property reads
\begin{equation}\label{smotion3}
  \theta_i(t+\Delta t) = \theta_i(t)+\frac{\Delta t}{k} \left[
    (\theta_j(t)-\theta_i(t)) \right.
  + \left. \cdots +( \theta_{j+k-1}(t)-\theta_i(t)) \right] .
\end{equation}
It is worth highlighting here that the very fact that relative states are
considered, prevents any singularity---such as those reported with the
original Vicsek's model~\cite{wei08:_singul}---from occuring. As already
mentioned, the formalism of Eq.~\eqref{smotion3} does not capture the fact
that the value of the $k$ indices---from $j$ to $j+k-1$ above---are actually
$i$-dependent since they are defined by the belonging, or not, of an agent to
the TNI of agent $i$. Moreover, these $k$ indices may change over time due to
the dynamical nature of the TNI, itself imposed by the dynamics of agent $i$.

The model devised here would not be realistic without accounting for the
ubiquitous presence of noise which may have disruptive behavioral
effects. This so-called behavioral noise can be divided into two broad
categories: the stimulus noise and the response
noise~\cite{dusenbery92:_sensor_ecolog}. The stimulus noise, a.k.a. intensity
noise, may have different origins like channel noise, environmental or
background noise, and receptor noise. In the present framework, the channel,
environmental and receptor noises are indistinguishable. In order to account
for the global effects of stimulus noise together with external perturbing
factors, a fixed level of background noise is considered throughout the
agents' surroundings. In addition, the response noise may have different
origins like motor noise and developmental noise which cannot be appropriately
included within the present idealized modeling framework. In what follows, the
response noise is therefore discarded and the stimulus noise may simply be
referred to as noise without any possible confusion.

Noise can generally be assumed to be random fluctuations with a normal
distribution~\cite{dusenbery92:_sensor_ecolog}. In the sequel, the background
noise is considered to have a normal distribution fully characterized by its
noise level, $\eta$. Specifically, the presence of noise modifies the equation
governing the dynamics of agent $i$ which now reads
\begin{equation}\label{smotion4}
  \theta_i(t+\Delta t) = \theta_i(t)+\frac{\Delta t}{k} \left[
    (\theta_j(t)-\theta_i(t)) + \cdots+ ( \theta_{j+k-1}(t)-\theta_i(t)) \right]+\Delta \theta,
\end{equation}
where $\Delta \theta$ is a random number chosen with a uniform probability
from the interval $[-\eta/2,\ \eta/2]$.

\subsection*{Simulation parameters}

%
In all simulations, agents are distributed across a 25--by--25 square with
periodic boundary conditions to avoid any boundary effect, while the time unit
$\Delta t=1$ was the time interval between two updates of the directions
$\theta_i(t)$ and the positions $\mathbf{x}_i(t)$ of each agent
$i=1,\cdots,N$. The synchronous position update is simply achieved through
\begin{equation}
  \mathbf{x}_i(t+\Delta t) =\mathbf{x}_i(t) + \mathbf{v}_i(t) \Delta t,
\end{equation}
where the velocity $\mathbf{v}_i(t)$ is calculated in its complex form $v_0
\exp(\text{i}\theta_i(t))$ with the constant speed $v_0$ taken equal to
0.05. Similarly to Vicsek \textit{et al.}~\cite{vicsek}, the value 0.05 for
$v_0$ was chosen such that agents always interact with their neighbors and
move fast enough to change the configuration after a few updates of the
directions. According to our simulations, in a wide range of the speed
($0.001<v_0<9$), the actual value of $v_0$ does not affect the results. In
most of our simulations, for the initial conditions, agents are initially
uniformly distributed in the two-dimensional spatial domain, with randomly
distributed directions.  Efficient ways of implementing such a swarm
simulation code are discussed in Ref.~\cite{ref:youseff,ref:lee,ref:helbing}.

The collaborative interactions of agents governs the dynamics of the
self-organization of the swarm, ultimately leading (or not) to the emergence
of consensus in the physical space. In the framework of our model, a good
metric for the consensus in the physical space is given by the average
alignment
\begin{equation}\label{eq:A1}
  \langle \vartheta \rangle = \frac{1}{N} \sum\limits^{N}_{i =
    1}\frac{v_i(t)}{v_0}=\frac{1}{N} \sum\limits^{N}_{i = 1}\exp \left(\textrm{i} \theta_i(t)  \right),
\end{equation}
over the $N$ agents of the swarm; $v_i(t)$ being the complex velocity of agent
$i$ in the plane at instant $t$. The alignment, $A$, is defined by the
absolute value of the steady-state average alignment: $A=|\langle \vartheta
\rangle (t_{\textrm{\tiny s}})|$, where $t_{\textrm{\tiny s}}$ is the time
required to reach a stationary state. This measure of the alignment approaches
the unity if all agents in the swarm move more or less in the same direction,
and is exactly equal to the unity if they are perfectly aligned. On the
contrary, if the agents fail to reach consensus, the alignment will tend to
zero, with the value $A=0$ representing utter mess.

\section*{Acknowledgments}

%
We thank Dr. Yilun Shang for fruitful and stimulating conversations. This work
was supported by a SUTD--MIT International Design Centre (IDC) Grant.

\end{document}